# Unveiling Mechanisms of SEI Formation and Sodium Loss in Sodium Batteries via Interface Reactor Sampling

Zhoulin Liu[abcd], Ziliang Wang[b*], Zherui Chen[e], Jianchun Sha[f], Fengzijun Pan[g], Pingyang Zhang[b],

Yinghe Zhang[acd*]

[a] School of Science, Harbin Institute of Technology, Shenzhen 518055, Guangdong, P.R. China

[b] National Engineering Laboratory for Reducing Emissions from Coal Combustion, Engineering Research Center of Environmental Thermal Technology of Ministry of Education, Shandong Key Laboratory of Green Thermal Power and Carbon Reduction, Shandong University, Jinan, Shandong 250061, P. R. China

[c] The Key Laboratory of Advanced Materials AI Manufacturing and Micro-Nano Processing, Harbin Institute of Technology, Shenzhen 518055, China

[d] Shenzhen Key Laboratory of Advanced Functional Carbon Materials Research and Comprehensive Application, Harbin Institute of Technology, Shenzhen 518055, China

[e] College of Applied Sciences, Shenzhen University, Shenzhen 518060, P.R. China

[f] Key Lab of Electromagnetic Processing of Materials, Ministry of Education, Northeastern University, Shenyang 110819, PR China

[g] School of Transportation Science and Engineering, Harbin Institute of Technology, Harbin 150090, P.R. China

*Corresponding authors. E-mail addresses: zwang2022@sdu.edu.cn (Ziliang Wang), zhangyinghe@hit.edu.cn (Yinghe Zhang)

**ABSTRACT:** The solid electrolyte interphase (SEI) critically dictates the cyclability and Coulombic efficiency of sodium-metal batteries, yet its dynamic formation mechanisms and atomic-scale evolution during electrochemical cycling remain elusive due to the spatiotemporal limitations of existing techniques. Here, an "Interface Reactor" sampling strategy is proposed to construct a charge-aware neuroevolution potential (qNEP). This approach overcomes the instability bottlenecks of conven-tional machine learning potentials, enabling stable, first-principles-accurate molecular dynamics simulations of complex electrode-electrolyte interfaces on the hundred-nanosecond scale. Fundamentally distinct SEI formation mechanisms are revealed during the early stage: carbonate-based electrolytes form heterogeneous organic-inorganic matrices via "mixed co-formation," whereas ether-based electrolytes generate dense, self-limiting inorganic barriers through "surface-energy-controlled" NaF crystallization. Metadynamics simulations further elucidate that these compositional disparities govern sodium-ion storage dynamics: NaF-rich SEIs facilitate efficient metallic deposition, while carbonate-dominated interphases induce irreversible sodium trapping and continuous electrolyte decomposition. These findings establish a comprehensive atomic-scale framework linking solvation structure, interfacial reaction networks, and electrochemical performance, providing mechanistic guidelines for rational SEI engineering in next-generation alkali-metal batteries. Crucially, a general and robust computational framework is established for simulating complex interfacial reactions in electrochemical systems.

## Introduction

The accelerating global transition toward renewable energy and electrified transportation has intensified the demand for sustainable and cost-effective energy storage technologies[1]. While lithium-ion batteries have dominated the market for decades, the geographically uneven distribution and rising cost of lithium resources motivate the exploration of alternative electrochemical systems[2]. In this context, sodium-based batteries have emerged as a promising candidate owing to the natural abundance, low cost, and favorable electrochemical properties of sodium[3]. Moreover, the chemical analogy between sodium and lithium offers the prospect of leveraging existing battery manufacturing infrastructure, rendering sodium batteries particularly attractive for large-scale grid storage and low-cost mobility applications[4,5]. Despite these advantages, the practical performance of sodium batteries remains critically constrained by interfacial phenomena at the electrode-electrolyte boundary[6,7]. During initial cycling, electrolyte decomposition leads to the formation of a solid electrolyte interphase (SEI) on the anode surface[8]. This nanometer-scale passivation layer must simultaneously suppress electronic transport while permitting rapid sodium-ion conduction[9–11]. As a result, the chemical composition, structural heterogeneity, and mechanical integrity of the SEI exert a decisive influence on interfacial stability, Coulombic efficiency, and long-term cycling performance[2,12]. Nevertheless, SEI formation remains one of the most critical yet least understood processes governing sodium battery operation.

Elucidating the formation and evolution of the SEI at the atomic scale poses a formidable challenge. Conventional experimental techniques, including X-ray photoelectron spectroscopy (XPS) and transmission electron microscopy, typically provide ex situ or ensemble-averaged information and struggle to capture the ultrafast, highly reactive interfacial processes occurring under operating conditions[8,13,14]. The intrinsic complexity of the interfacial environment, coupled with the rapid and irreversible nature of electrolyte decomposition, has thus far precluded a unified mechanistic picture of SEI formation[8]. Consequently, fundamental questions concerning reaction pathways, charge-transfer processes, and long-term SEI evolution remain unresolved, hindering the rational design of high-performance sodium battery interfaces[2].

Atomic-scale simulations offer a powerful complementary approach for probing interfacial chemistry beyond current experimental resolution. Ab initio molecular dynamics (AIMD) has provided valuable insights into elementary reaction mechanisms, but its high computational cost confines simulations to small systems and short time-scales, typically only hundreds of atoms and tens of picoseconds[15,16]. To extend simulations to larger length and time scales, reactive molecular dynamics, particularly reactive force field such as ReaxFF, has been widely applied to investigate SEI formation[5,17–19]. However, the predefined functional form and reaction-specific parameterization of conventional ReaxFF fundamentally limit its transferability and accuracy for complex electrochemical interfaces[5,18,19]. Notably, ReaxFF simulations of carbonate electrolyte decomposition at alkali-metal electrodes have failed to capture essential single-electron reduction pathways, leading to qualitatively incorrect product distributions[18,20]; these deficiencies were later resolved only by high-accuracy ab initio and machine learning molecular dynamics approaches[16,21].

Machine learning potentials (MLPs) have recently emerged as a transformative framework for modeling interatomic interactions with near first-principles accuracy at substantially reduced computational cost[22,23]. Importantly, MLPs have already demonstrated remarkable success across a wide range of battery-related systems beyond electrode-electrolyte interfacial chemistry, including bulk electrolytes[24], solvation structures, ion transport, and solid-phase products, where nanosecond or longer simulations have been achieved while maintaining high fidelity[25–27]. In these relatively homogeneous environments, MLP-based molecular dynamics has enabled detailed investigations of structural evolution and phase stability that are inaccessible to conventional quantum mechanical methods.

In stark contrast, extending such success to electrode-electrolyte interfaces has proven far more elusive[28–30]. Interfacial systems simultaneously involve heterogeneous local chemical environments, complex bond-breaking and bond-forming reactions, strong electric fields, and continuous charge transfer across the interface[28]. These coupled effects generate an exceptionally rugged and high-dimensional potential energy surface, rendering interface simulations particularly susceptible to instability and error accumulation during long-timescale molecular dynamics[29,31,32]. As a result, existing MLP-based metal electrode-electrolyte simulations remain limited to ultrashort durations ($\sim 100$ ps), insufficient to capture the kinetic evolution of interfacial structures[21,31,32]. Crucially, the requirement for long-timescale simulations is not merely a computational aspiration but a fundamental physical necessity. Key interfacial phenomena governing SEI maturation, such as the densification[33] and phase separation of inorganic components, intrinsically occur on nanosecond or longer timescales[28,34,35]. Simulations confined to picosecond or sub-nanosecond regimes are therefore intrinsically incapable of resolving the mechanisms that control SEI growth, stability, and its feedback on metal deposition behavior. Achieving stable, chemically accurate, and long-timescale molecular dynamics at electrode-electrolyte interfaces thus represents a central methodological bottleneck in computational battery research.

Here, we introduce an Interface Reactor sampling strategy to construct a charge-aware neuroevolution potential (qNEP) specifically designed for metal electrode–organic electrolyte interfaces. Guid-ed by high-temperature AIMD sampling of interfacial reactions, this framework achieves a robust balance between accuracy and efficiency, extending simulation stability by two to three orders of magnitude relative to existing approaches. The resulting qNEP enables first-principles-precision molecular dynamics simulations of interfacial reaction networks in systems containing tens of thousands of atoms over a hundred nanoseconds, allowing direct atomistic observation of the complete SEI formation process. By explicitly incorporating trainable atomic charges, the model captures charge-transfer dynamics during electrolyte decomposition and sodium-ion migration, providing new mechanistic insights into interfacial storage and transport. Using this approach, we reveal fundamentally distinct SEI formation and growth mechanisms in carbonate- versus ether-based electrolytes and, through metadynamics simulations, uncover the atomic-scale evolution of the SEI during charging and its decisive role in regulating sodium deposition morphology and irreversible sodium loss. Together, this work establishes a comprehensive atomistic framework linking solvation structure, interfacial reactivity, and electrode storage behavior, offering a transferable approach for investigating complex electrochemical interfaces.

## Methods

**The qNEP training hyperparameters.** Graphics Processing Units Molecular Dynamics (GPUMD) v4.3 was used to train the qNEP model[36,37]. Detailed information regarding the multi-loss training algorithm based on the qNEP model and SNES, as well as the nep.in input file used, can be found in Supplementary Note S1–S2.

In general, the selection of the cutoff radius and the number of neurons presents a trade-off among computational speed, accuracy, and the risk of overfitting[26]. In contrast, the choice of $n\_max$, $l\_max$, and $batch$ size primarily involves a trade-off between accuracy and speed[38]. Therefore, to balance these competing factors, we prioritized increasing the number of $neurons$, $n\_max$, and the $batch$ size. The NEP-4 model[26] was trained with the batch size $N_{bat}$ of full batch, while the number of neurons $N_{neu}$ was set to 50. The radial descriptor indices $n_{max}^{R}$ of 8 and angular descriptor indices $n_{max}^{A}$ were 8. The numbers of radial and angular descriptor basis functions, $N_{bas}^{R}$ and $N_{bas}^{A}$ were both set to 8. Concurrently, a radial cutoff of 6.5 Å and an angular cutoff of 4.0 Å were set to achieve an optimal balance between accuracy and resolution. The population size $N_{bat}$ was set to 50, and steps $N_{gen}$ was set to 500,000. The training took approximately 2 days on $4 \times$ RTX 4090 GPUs. The total number of trainable parameters in the model for 6 elements is 22,633, which demonstrates the data efficiency of the NEP model[39,40].

**Machine Learning Molecular Dynamics (MLMD)** After completing the training of the MLP, the MLMD simulations were performed. Before the simulations, the temperature was set to 300 K and the pressure to 1 atm. The system was then relaxed under an isothermal-isobaric ensemble (NPT) to reduce internal stresses. Subsequently, the system was heated to the desired temperature and simulated using the Langevin thermostat in the canonical (NVT) ensemble.[41] The time step was set to 0.5 fs. To balance computational efficiency, chemical reaction resolution, and the size of the output files, the total simulation time was set to 0.4–100 ns. All MD simulations can be performed using a single RTX 4090 GPU, with the longest simulation taking approximately 8 days.

**DFT Calculation Details.** The electronic structure calculations were performed using the Vienna Ab initio Simulation Package (VASP).[41] The Perdew-Burke-Ernzerhof (PBE)[42] generalized gradient approximation (GGA) was employed to describe the exchange-correlation energy, which is widely used in similar studies[32,43]. To account for core–valence interactions, the projector augmented-wave (PAW) method[44] was employed with the energy cutoff of 600 eV. For Brillouin zone sampling, a Monkhorst-Pack k-point grid centered[45] at the Gamma point was used, with a k-point sampling interval of 0.04 Å$^{-1}$ for periodic systems. Considering the multiple components in the system, the electronic level occupation was treated with a Gaussian smearing[46] of 0.05 eV. Entropy corrections due to thermal occupation were subtracted from total energy calculations. The Grimme D3 dispersion correction was applied to account for long-range dispersion interactions[47]. The convergence criterion for electronic self-consistency interactions was set to $1 \times 10^{-6}$ eV·unit$^{-1}$. The calculations are automated using the NepTrain software[48]. The complete VASP input parameters for static calculations are summarized in Supplementary Note S3. Net force analysis confirmed full convergence of the entire training set (Supplementary Note S4).

**Results and Discussion**

**The Interface Reactor Methodology.** Simulating electrode-electrolyte interface systems using empirical MD or AIMD directly presents multiple challenges due to their inherent structural and compositional complexity[32]. The interfacial potential energy surface is highly rugged, so the phase-space distribution can drift readily, which reduces the probability of sampling reactive configurations. In addition, molecules in the interfacial layer adopt strongly anisotropic orientations, and their conformational relaxation is often incomplete within accessible simulation times. If the underlying potential is even slightly biased, the system may follow unphysical energy pathways, causing numerical instability or simulation failure. Accordingly, a high-accuracy machine-learning potential for reactions at metal electrode–organic electrolyte interfaces requires a training set that is as exhaustive as possible in descriptor space, configuration space and chemical-composition space.

First, regarding the descriptor space, a high-dimensional neural network potential method was employed based on atom-centered symmetry functions. As illustrated in Figure 1a,b the local atomic environment is characterized by a set of radial and angular descriptors. Radial descriptors capture the spatial distribution of neighboring atoms within a given cutoff radius, while angular descriptors, through many-body correlations and spherical harmonics, describe the relative orientations between atoms (the complete mathematical definitions and explicit expressions for both types of descriptors are provided in Supplementary Note S5). The descriptor space plots for the AIMD results used to prepare the initial training set are shown in Figure 1c. As the sampling temperature increases, the coverage of descriptors in the high-dimensional space gradually expands. The descriptor set obtained at high temperatures nearly completely encompasses the distribution observed at lower temperatures. Therefore, generating the initial training set via high-temperature MD can significantly enhance the ergodicity of local atomic environments within the descriptor space.

Second, during the iterative expansion of the configuration space, if one relies entirely on MLP-driven MD trajectories, the incompleteness of the potential can readily cause the system to deviate from physically reasonable regions, becoming a major source of unphysical configurations. As shown in Figure 1d, in certain regions, although the MLP predicts a low energy, the true DFT energy is significantly higher than that of normal configurations. This indicates that these regions corre-spond to high-energy states erroneously "smoothed out" by the MLP, stemming from the misjudgment of key energy barriers. The local potential energy surface illustrated in Figure 1e further demonstrates that MLMD trajectories may select incorrect paths near energy barriers and gradually diverge from the true reaction channels. To suppress such trajectory bifurcation and maintain the physical relevance of configuration space exploration, short-range AIMD simulations are introduced to correct critical regions just before the MLMD trajectory approaches a bifurcation point. This imposes "physical guidance" on the trajectory, steering it back to evolve along reaction pathways with lower true energy barriers that align with chemical intuition. The introduction of this physical constraint during the active learning process effectively prevents the accumulation of unphysical configurations and the systematic drift of the potential.

Third, for the MLPs to reliably describe actual reactions at the metal electrode–organic electrolyte interface, their training set must also achieve sufficient coverage in terms of chemical composition, encompassing potential intermediates and products that may arise in the substrate, interfacial region, and bulk solution phases. Specifically, during dataset construction, representative inorganic products (e.g., NaF, $Na_2O$) likely generated from interfacial reactions, as well as various configurations of organic decomposition intermediates in the solution phase were explicitly incorporated. Their distinct local environments near the substrate, within the interfacial layer, and in the bulk electrolyte were concurrently explained. By systematically including these potential intermediates and products into the active learning loop, the training set achieves more comprehensive coverage across both reaction pathways and final-state spaces.

In summary, by utilizing high-temperature sampling to achieve broad traversal of the descriptor space, introducing AIMD-guided physical corrections at critical nodes during active learning to constrain configuration space exploration, and systematically traversing the potential intermediate and product configurations from interfacial reactions, we have constructed a complete data-generation loop. This loop consists of an initial training set, reaction-path correction, and product/intermediate complementation. This approach ensures the accuracy and stability of the MLPs for complex electrode-electrolyte interfacial reactions.

Based on the aforementioned approach, the entire MLP training process is divided into two stages: initial MLP training and active learning iteration. During the initial potential training stage, structures were obtained through multiple methods, including perturbation, AIMD simulations, and sampling with the NEP89 potential[49]. Their energies, forces, and virials were then computed using DFT methods. The active learning process is illustrated in Figure 1f. After the initial potential was constructed, structures were generated via MD simulation and sampled to optimize the training set. Crucially, intensive AIMD sampling was introduced to guide the trajectory back to a physically sound path just before the emergence of unphysical configurations that would cause simulation failure. The sampled results were then labeled with DFT calculations and added to the training set, followed by subsequent NEP training. This cycle was repeated until the system could stably run for 100 ns. Detailed protocols and tools[48,50] for the initial training set sampling and the rules for physics-guided correction can be found in Supplementary Note S6. The final training set comprises 5035 configurations containing a total of 683,526 atoms; its detailed composition is provided in Supplementary Note S7.

Finally, as shown in Figure 1g, during a 100 ns MLMD simulation of a Na electrode-electrolyte system containing ~1000 atoms, the potential remained stable throughout. Compared to similar systems (Li/Na electrode-electrolyte) reported in references[21,31,32], this represents a stability improvement of approximately 2–3 orders of magnitude. For a larger system containing ~27,000 atoms, simulations on the scale of tens of nanoseconds were also achievable (see Supplementary Note S8). Therefore, these results validate the success of the "Interface Reactor" sampling methodology for metal electrode-electrolyte simulations. It enables effective in situ observation of interfacial reactions at DFT-level accuracy, thereby bridging the gap between microscopic structure and macroscopic performance.

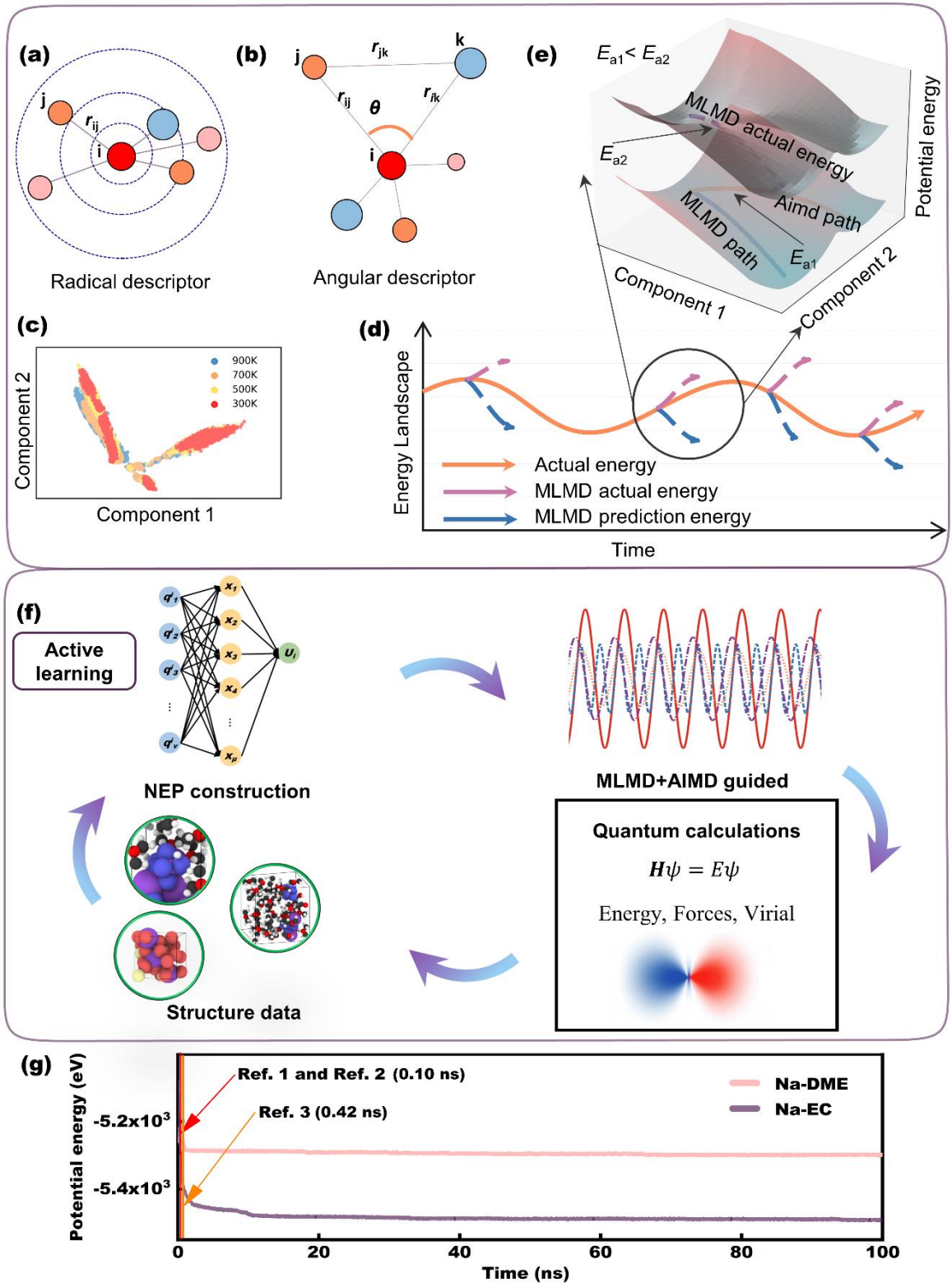

Figure 1. (a) Radial descriptors; (b) Angular descriptors; (c) Relationship between temperature and descriptor components; (d) Schematic of AIMD-guided configuration space exploration; (e) Schematic of a local potential energy surface; (f) Flowchart of the active learning process; (g) Comparison of the achievable simulation timescales for the trained NEP-MLP with those reported in the literature.

**Performance of the Trained MLP Model.** As shown in Figure 2, the accuracy, computational speed, and charge characteristics of the NEP-MLP model are investigated in detail. Figure 2a presents the loss function curves, demonstrating that all parameters converged after 500,000 iterations. The parity plots for energy and forces (Figure 2b,c) confirm the high accuracy of the NEP model. Typically, multi-element systems and coexisting solid/liquid phases can degrade fitting accuracy, yet the NEP model maintains excellent performance under these conditions. To reduce the computational cost of evaluating the test set, structures remaining after constructing the training set were used as the test set, which comprises 1076 configurations containing a total of 268,641 atoms. The root mean square errors (RMSEs) for the training set are 10.50 meV·atom$^{-1}$ for energy and 251.89 meV·Å$^{-1}$ for forces. The corresponding errors for the test set are 12.11 meV·atom$^{-1}$ and 250.36 meV·Å$^{-1}$, respectively.

In MLPs, descriptors are key components that transform the local atomic environment into mathematical vectors to capture essential structural features. The descriptor vectors in NEP training consist of multiple radial and angular components, where the radial functions are defined as linear combinations of basis functions[26]. These descriptor components are invariant to permutations of atoms of the same type, and their coefficients act as trainable parameters, which is crucial for effectively distinguishing the contributions of different atom pairs within the descriptor. As the training data generation method relies on the chemical transferability embedded in the radial function equations, we can validate this feature through principal component analysis (PCA) of the descriptor space (Figure 1d). The test set is fully encompassed by the training set and exhibits a continuous, contiguous distribution rather than discrete, island-like clusters, indicating that the training set provides remarkably comprehensive coverage of the configuration space for the elements under study. Furthermore, results of 1 ns sampling test also demonstrate the high accuracy of the potential (Figure 1e,f).

To further verify the model's accuracy, a ReaxFF commonly used to describe electrolyte solutions and Li/Na reactions[51] was selected for comparison with our model. The forces for the training set were calculated using ReaxFF and compared with the results from the trained NEP-MLP (Supplementary Note S9). The results show that the RMSE of NEP-MLP is 251.89 meV/Å, significantly lower than ReaxFF's RMSE of 9680 meV/Å, indicating that NEP-MLP accuracy substantially surpasses that of traditional ReaxFFs. Additionally, NEP-MLP delivers exceptional computational speed in MD simulations. On a single RTX 4090 GPU, its speed for large systems is approximately one order of magnitude faster than traditional ReaxFF methods and about six orders of magnitude faster than ab initio precision AIMD methods (Figure 2g). Therefore, it readily enables million-atom simulations, thereby forming the foundation for our discoveries. Moreover, standard MLPs often lack the explicit modeling of charge/valence states and their associated long-range electrostatic interactions, which typically limits the observation of chemical reactions. However, our qNEP model, which incorporates charge information, addresses this shortcoming. This model successfully learns the relationship between atomic environment and charge distribution, effectively distinguishing atoms in different chemical states (Figure 2h). This capability will readily allow us to access charge-transfer-related chemical properties, such as the charge transfer of sodium ions at the interface and their storage modes within the electrode. Further validation of EC ring-opening barriers and sodium melting point confirms model accuracy (Supplementary Notes S10–S11). These validations confirm the model's reliability, efficiency, and its capability to resolve charges in complex chemical processes.

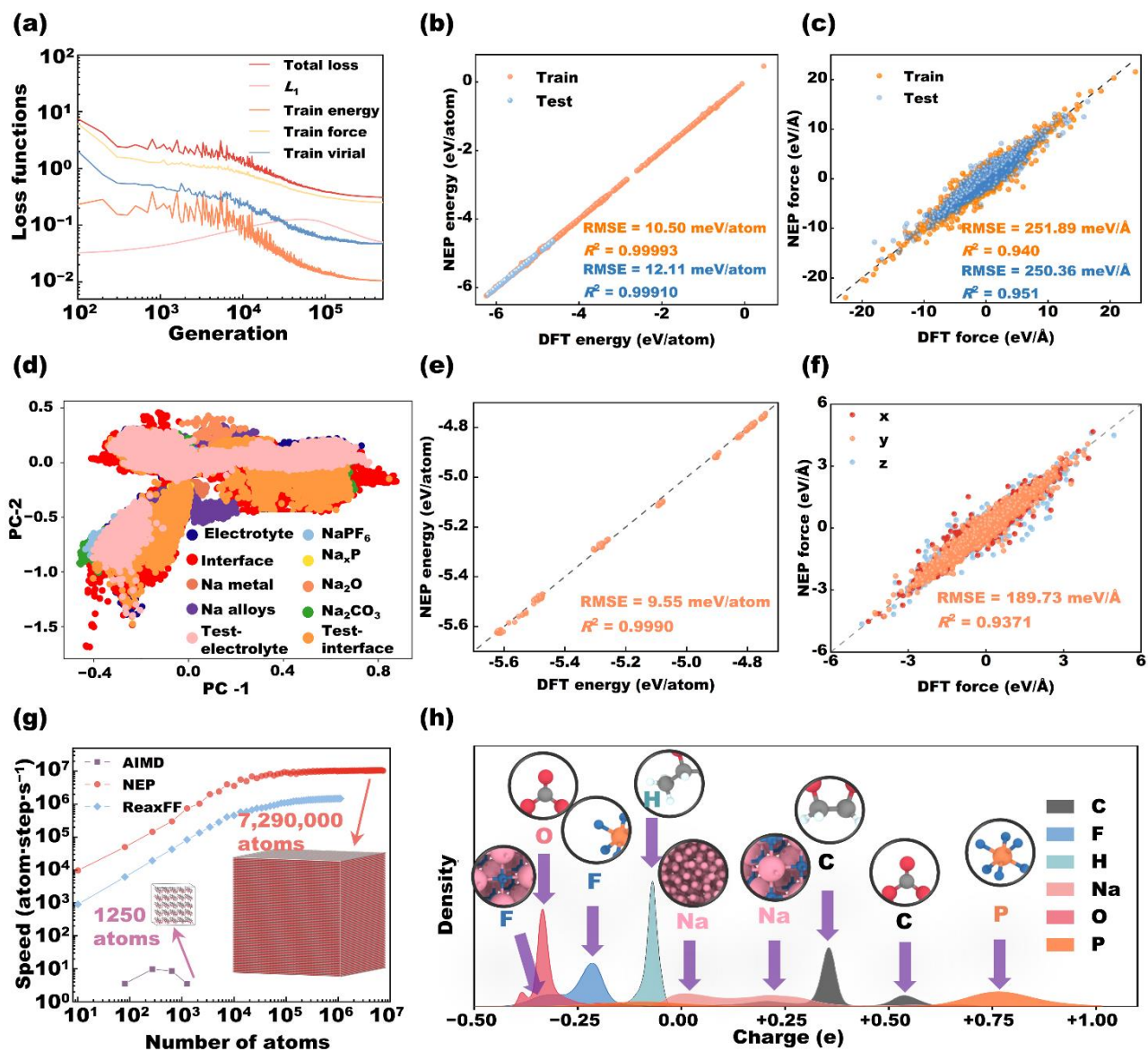

Figure 2. (a) Loss function during NEP-MLP training; (b) Energy parity plot for the NEP model; (c) Force parity plot; (d) PCA projection of the training set; (e) 1 ns sampling test for energy; (f) 1 ns sampling test for forces; (g) Comparison of computational speed between AIMD, ReaxFF-MD, and NEP-MD; (h) Predicted charges for the training set and their corresponding atomic environments.

**Distinct Decomposition Pathways of Carbonate and Ether Electrolytes.** Elucidating the decomposition pathways of electrolytes at the electrode surface provides fundamental physicochemical insights for understanding SEI formation mechanisms and guiding electrolyte and additive design[2]. Traditionally, constrained by time and resource costs, investigating the chemical reactions of a single electrolyte has been a significant undertaking[16,52]. Owing to the high efficiency and transferability of our NEP-MLP, we can conveniently utilize MLPs to study the general reaction patterns of these two electrolyte classes. Five common sodium-metal battery electrolytes, including three carbonates (eth-ylene carbonate EC, dimethyl carbonate DMC, propylene carbonate PC) and two ethers (dimethoxy-ethane DME, diglyme G2), were selected to explore their interfacial reaction behaviors with a sodium metal electrode via MLMD simulations. We focus our detailed analysis on two representative systems, EC and DME; detailed comparisons for the other electrolytes are provided in Supplementary Note S12.

Figure 3a,b shows the initial structures of 1 M $NaPF_6$ in EC and DME, respectively. The EC system exhibits extremely high reactivity at both 350 K and 450 K (Figure 3c,d). Its decomposition primarily follows two characteristic pathways, which are similar to those reported for Li[16] and Ca[52]. The first path involves a two-electron transfer, featuring cleavage of the ester C=O double bond to generate carbonate species remaining on the electrode surface while releasing ethylene gas into the solution. The other path is a single-electron transfer process, where two C–O single bonds within the carbonate ring break to form disodium ethylene glycolate ($NaO–CH_2–CH_2–ONa$) and CO. Furthermore, at 450 K, this process nears completion within approximately 100 ps, indicating that the interfacial decomposition of carbonates is highly rapid.

In contrast, the DME system shows significantly lower reactivity (Figure 3e,f). At 350 K, only the slow scission of a single C–O bond, generating methyl radicals, is observed, with no gas evolution. Only when the temperature is increased to 450 K are small amounts of ethylene and ethane gas products formed, indicating that the interfacial decomposition capability of ether-based solvents is much weaker than that of carbonates.

Notably, in both solvent types, the decomposition of $PF_6^-$ occurs relatively quickly and shows no strong dependence on the solvent. Its reaction pathway can be summarized as a reduction reaction with sodium, forming NaF and phosphorus-containing compounds. These observations strongly suggest that the surface SEI growth mechanism is governed by the reactivity of the solvents/additives.

Species evolution diagrams (Figure 3g–i) further highlight the mechanistic differences between the two types of systems. In the EC system, the pathway leading to carbonate formation accompanied by ethylene release is dominant, while two competing pathways yielding organic sodium compounds also occur, including a minor reaction generating $CO_2$. In contrast, the DME system exhibits a more prominent pathway generating organic sodium compounds, albeit far less significant than the decomposition of $NaPF_6$.

A systematic comparison of carbonate electrolytes (EC, PC, DMC) with ether-based solvents (DME, G2) (Supplementary Note S12, Figure 3j–o) clearly reveals high intra-class consistency across bond cleavage sites (Figure 3j,k), evolution of reaction intermediates (Figure 3l,m), and final products (Figure 3n,o). Carbonates consistently tend to produce carbonates, organic sodium salts, and gaseous hydrocarbons, reflecting their strong interfacial reactivity. Ether solvents primarily yield NaF, alkanes, and sodium alkoxides, indicating overall weaker reactivity.

Therefore, MLMD simulations reveal a general reaction pattern: carbonate-based solvents tend to decompose rapidly, forming an SEI rich in organic components, while ether-based solvents exhibit lower reactivity, leading to rapid NaF formation. There are fundamental differences in decomposition pathways and product composition between carbonate-based and ether-based electrolytes. This strongly indicates that the mechanisms unveiled are universal rather than specific to individual systems.

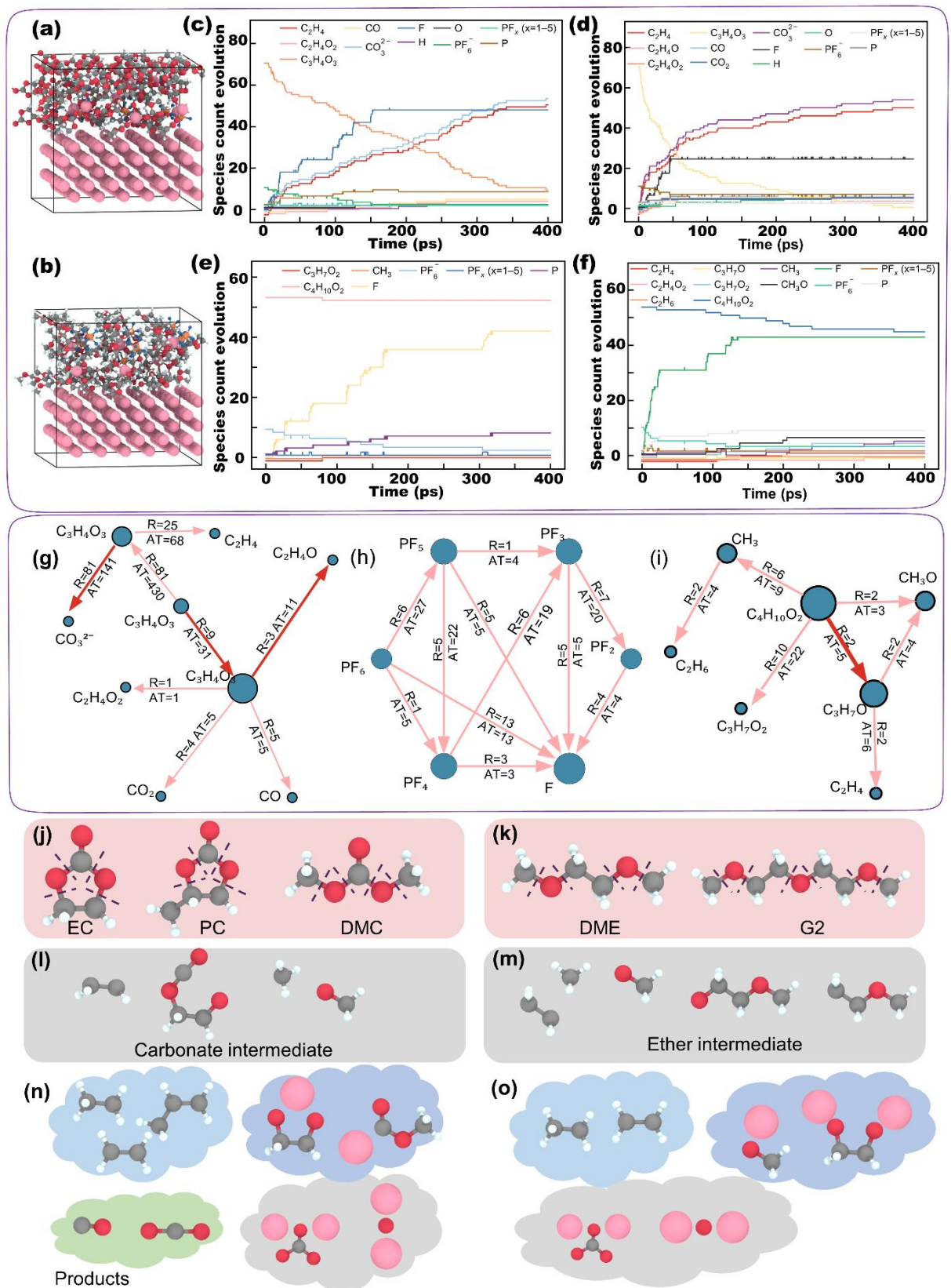

Figure 3. (a) Schematic of the model for 1M NaPF₆ in EC on a Na electrode; (b) Schematic of the model for 1M NaPF₆ in DME on a Na electrode. Species evolution diagrams: (c) Na-EC system at 350 K; (d) Na-EC system at 450 K; (e) Na-DME system at 350 K; (f) Na-DME system at 450 K (some curves are slightly offset to prevent overlap). Species transport diagrams at 450 K: (g) Na-EC

system; (h) NaPF$_6$; (i) Na-DME system. Schematic of bond cleavage sites for reactions with sodium in: (j) carbonate-based solvents; (k) ether-based solvents. Schematic of reaction intermediates for reactions with sodium in: (l) carbonate-based solvents; (m) ether-based solvents. Schematic of reaction products for reactions with sodium in: (n) carbonate-based solvents; (o) ether-based solvents. Atoms are represented in a ball-and-stick model (C: grey, F: blue, H: white, Na: pink, O: red, P: orange).

**SEI Growth Patterns Governed by Relative Reactivity and NaF Growth Characteristics.** To delve into the dynamic formation mechanism of the SEI, further MLMD simulations were conducted at extended timescales (100 ns) and with large system sizes (~27,000 atoms). Snapshots from smaller-scale simulations (Figure 4a–d) clearly illustrate the significant kinetic differences between EC and DME electrolytes. In the EC system, the reactions are largely complete within the initial 10 ns, yielding products such as sodium carbonate, ethylene, and organic sodium salts. In contrast, for the DME system, even after 100 ns of simulation, most of the metallic sodium and electrolyte show no significant reaction, with NaF being the primary product.

This distinction is more profoundly revealed in the large-system simulations (Figure 4e–j). In the EC electrolyte, various decomposition products (e.g., Na$_2$CO$_3$, organic sodium salts) nucleate and grow rapidly and simultaneously at different sites on the electrode interface, forming an SEI with a mixed organic-inorganic co-growth structure (Figure 4i). Notably, gaseous products released from reactions (e.g., C$_2$H$_4$) may block electrolyte-electrode contact. Conversely, in the DME electrolyte, the formation rate of NaF far exceeds that of other by-products. During the initial growth stage, F$^-$ ions released from PF$_6^-$ decomposition randomly dissolve into the sodium matrix. Subsequently, disordered sodium fluoride aggregates and undergoes heterogeneous nucleation, ultimately growing on the sodium substrate in a spherical cap morphology (Figure 4d,g). As illustrated in Supplementary Note S13, the (001) facet of NaF possesses the lowest surface energy. Following sufficient fluoride aggregation (from stage 1 to stage 2 in Figure 4g), the amorphous phase transitions into an ordered crystal. To minimize surface energy, the NaF crystal grows outward from the underlying sodium metal, resulting in a morphology consistent with the calculated Wulff construction (Figure 4h).

This process aligns with classical nucleation theory[53] and is a typical heterogeneous nucleation process governed by surface-energy-controlled growth[54,55]. This mechanism promotes the preferential lateral extension of NaF, forming an ordered, dense, sheet-like overlayer (Figure 4e), which subsequently dictates the SEI's composition and morphology, resulting in a denser SEI layer primarily composed of inorganic species at the early stage (Figure 4j).

The phenomena described above stem from the combined effects of the relative decomposition rates of electrolyte components and the unique surface-controlled growth mode of NaF on the Na electrode. As revealed in Figure 3, carbonate-based solvents exhibit high decomposition activity. Consequently, this high solvent reactivity drives the rapid co-deposition of multiple products, forming a mixed structure. In the DME system, the preferential decomposition of the salt makes NaF the dominant product. Its surface-energy-controlled growth mechanism further determines the final SEI morphology: this mechanism promotes the lateral spreading of NaF, creating a dense physical barrier. This overlayer not only achieves self-limiting growth but also significantly inhibits the subsequent decomposition of solvent molecules, thereby kinetically locking in the inorganic-dominated nature of the SEI.

Two typical SEI formation mechanisms are identified at the atomic scale: the EC electrolyte corresponds to a "mixed co-formation mechanism", whereas the DME electrolyte follows a "surface-energy-controlled preferential growth mechanism". This indicates that the relative reactivity among electrolyte components, by directing the chemical composition of the initial products, triggers fundamentally distinct growth patterns. This understanding may serve as a key to tailoring SEI structure and properties.

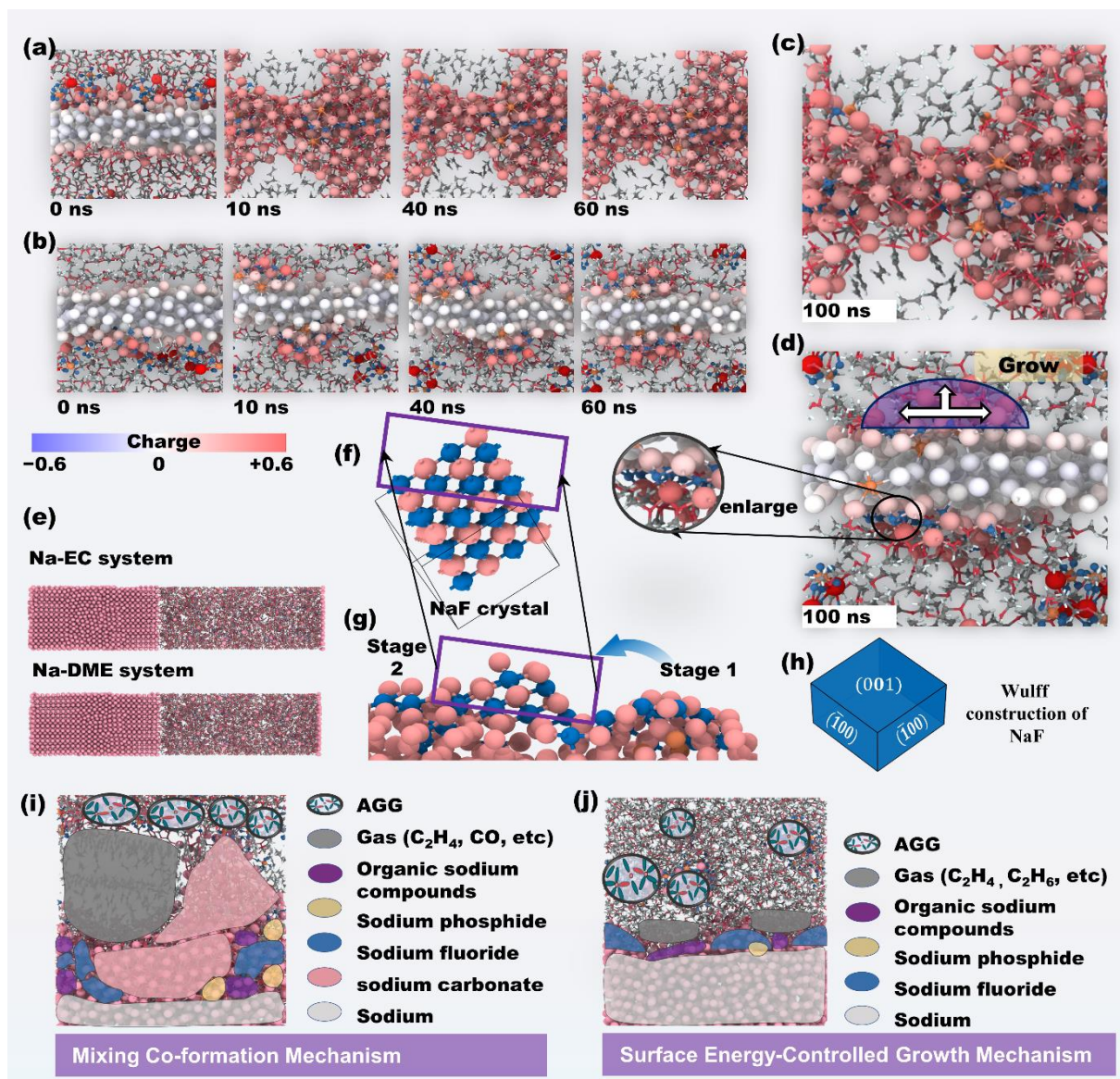

Figure 4. Small-system results at 350 K: (a) Reaction snapshots of the Na-EC system from 0–60 ns; (b) Reaction snapshots of the Na-DME system from 0–60 ns; (c) Reaction snapshot of the Na-EC system at 100 ns; (d) Reaction snapshot of the Na-DME system at 100 ns; Large-system results at 350 K: (e) Models of the Na-EC system and Na-DME system; (f) NaF crystal; (g) NaF crystal growth in the large system; (h) Wulff construction of NaF; (i) Results of the Na-EC system at 40 ns; (j) Results of the Na-DME system at 40 ns. Atoms are represented in a ball-and-stick model (C: grey, F: blue, H: white, Na: pink, O: red, P: orange); within the small system, Na atoms are colored according to their partial charges.

**Experimental Corroboration by Gas Chromatography (GC) and XPS.** To validate the decomposition pathways, GC and XPS analyses were performed on sodium anodes immersed in two distinct electrolytes. The experimental results are highly consistent with atomic-scale predictions. First, gas analysis (Fig. 5a,b) verifies distinct reaction channels. In EC, $C_2H_4$ (58.6%) validates the two-electron carbonate pathway (Fig. 3g). In contrast, DME gas is dominated by $C_2H_6$ (42.0%) and $CH_4$ (18.7%), reflecting C–O cleavage and recombination (Fig. 3i). Temperature dependence (Fig. 5b) shows increased $CO_2$. Theoretically, PBE function calculations predict CO is more favorable than $CO_2$ ($\Delta G_{CO} < \Delta G_{CO_2}$, Supplementary Note S**14**), aligning with MLMD (Supplementary Note S15). However, experiments show higher $CO_2$ (Fig. 5b). This discrepancy likely stems from the inherent limitations of the PBE functional, yet it can be rationalized by thermodynamic principles. With its significantly higher standard entropy compared to CO (214.3 versus 197.8 J/(mol·K) for $CO_2$ and CO, respectively), the formation of $CO_2$ becomes increasingly favored at elevated temperatures as the Gibbs free energy change, $\Delta G$, becomes more negative. Second, XPS quantitative analysis (Fig. 5c,d) validates SEI origins. High $Na_2CO_3$ in EC (6.9% vs 1.36%) confirms carbonate reactivity. In DME, higher organic sodium (5.08% vs 3.92%) and NaF (1.44%) validate solvent alkylation and salt decomposition. Collectively, these experiments validate qNEP reliability and bridge atomic mechanisms with macroscopic composition.

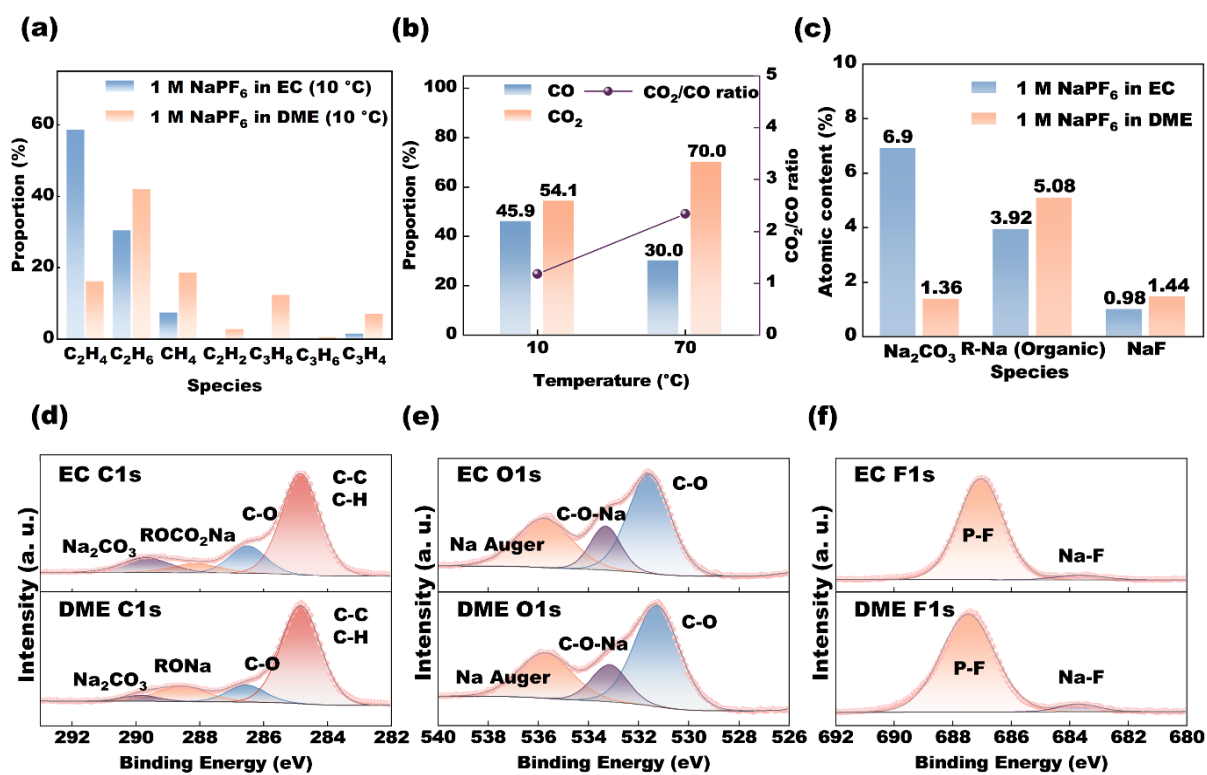

Figure 5. Experimental validation of interfacial decomposition products. (a) Gaseous product composition in distinct electrolytes; (b) Temperature dependence of $CO_2/CO$ ratios in EC system; (c)

Quantitative species content of SEI components in EC and DME systems; (d) C1s XPS spectra; (e) O1s XPS spectra; (f) F1s XPS spectra.

**Regulation of SEI Composition by Na$^+$ Solvation Structure.** The solvation structure of sodium ions in the electrolyte is a key microscopic factor governing interfacial reactions and SEI formation. Na$^+$ exists primarily in three forms within the electrolyte: solvent-separated ion pairs (SSIPs), contact ion pairs (CIPs), and aggregates (AGGs)[56]. Quantitative analysis of radial distribution functions (RDFs) and coordination numbers (CNs) reveals that electrolyte concentration and solvent chemistry critically modulate the solvation environment, dictating the species that ultimately reach the interface.

As shown in Figure 6a, increasing the NaPF$_6$ concentration from 0.5 M to 2 M significantly alters the solvation structure. The intensity of the main Na–O peak (Figure 6c) and coordination numbers (Figure 6d) decrease, confirming the competition between PF$_6^-$ anions and solvent molecules within the solvation shell. This shifts the balance from SSIPs to CIPs and AGGs (Figure 6b and Supplementary Note S16). Detailed analysis of temperature effects on solvation dynamics is provided in Supplementary Note S17. Furthermore, solvent type determines the distinct solvation shell configurations; carbonate-based electrolytes exhibit sharper, more ordered solvation shells compared to the more flexible ether-based solvents, as reflected in the CNs (Figure 6c,d).

The aforementioned characteristics of the solvation structure directly govern the SEI formation mechanism. The bulk solvation structure and the proportions of SSIPs, CIPs, and AGGs essentially determine the desolvation energy barrier at the interface and the resulting SEI composition. While SSIPs and CIPs possess high mobility, AGGs exhibit significantly poorer mobility[57]. Consistent with the interfacial SEI growth shown in Figure 4i–j, SSIPs and CIPs are nearly absent in the immediate interfacial region upon SEI formation, leaving behind the less mobile AGGs. Furthermore, once AGGs come into contact with the interface, they can readily form surface protrusions that act as sites for non-uniform SEI growth (Supplementary Note S18). Consequently, the electrolyte composition, by regulating the bulk solvation structure, dictates the identity and reduction sequence of precursor species arriving at the interface. This hierarchical regulation from composition to structure and inter-facial reactions governs the composition and growth patterns of the SEI, providing insight into the differences in SEI formation across different electrolytes.

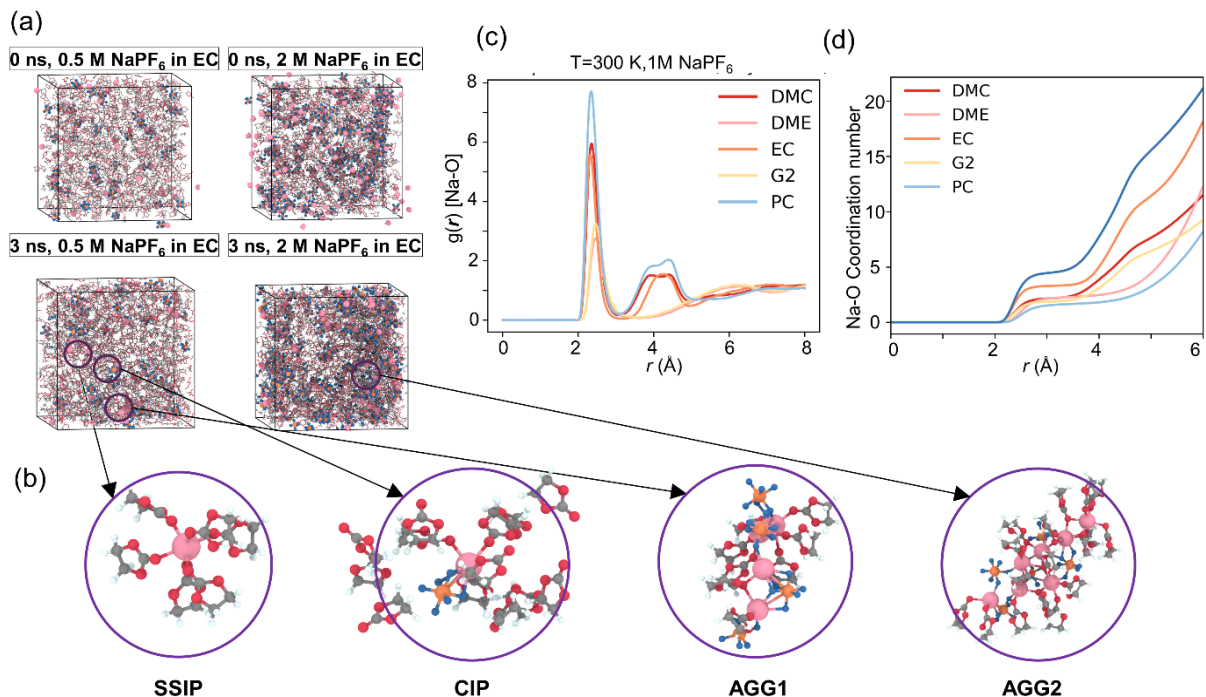

Figure 6. MD simulations of EC electrolytes: (a) Initial models and MD snapshots (3 ns) for 0.5 M and 2 M NaPF$_6$ in EC systems; (b) Schematic of sodium ion solvation structures; (c) RDFs for 1 M NaPF$_6$ in EC systems at 300 K; (d) Na–O Coordination numbers (CNs) for different electrolytes; Atoms are represented in a ball-and-stick model (C: grey, F: blue, H: white, Na: pink, O: red, P: orange).

**Dynamic Evolution of the SEI and Sodium Loss.** The electrode interface typically comprises the electrode, the electrolyte, and the SEI formed by their interaction, as illustrated in Figure 7a. Understanding the dynamic storage behavior of sodium ions at this interface during charge/discharge is key to unveiling its evolution mechanism. Specifically, investigating the processes of sodium ion storage and release is crucial for comprehending the details of this evolution. To reveal the dynamic behavior and storage mechanism of sodium ions crossing the electrode-electrolyte interface during cycling at the atomic scale, we employed metadynamics simulations. This method effectively overcomes the limitations of conventional MD in simulating sodium storage processes.

During the charging process (Figure 7b), by continuously applying a bias potential to the potential energy surface, the metadynamics simulation captures the dynamic process of sodium ions being ex-tracted from the bulk electrolyte, traversing the interface, and ultimately being deposited into the anode. The sodium deposition rate was set to a constant 0.1 ns/atom (simulation details are provided in Supplementary Note S19). We focused on interfaces featuring two typical SEI components: $Na_2CO_3$ and NaF.

As illustrated by the interface models in Figure 7c,e, the system consists of the electrolyte, SEI layer, and underlying anode substrate (graphene). During charging, the SEI layer thickens in both systems as the upper electrolyte is consumed (Figure 7c,e, and Supplementary Note S20). However, they exhibit significant differences in reactivity and deposition behavior. In the $Na_2CO_3$-dominated system, the electrolyte is nearly depleted, indicating severe side reactions. The sodium stored in the substrate appears light pink (Figure 7d), suggesting ionic character. In contrast, in the NaF-dominated system, electrolyte side reactions are effectively suppressed. The deposited sodium appears pure white (Figure 7f), indicative of metallic character, and the amount of stored sodium is significantly greater.

Quantitative analysis further elucidates these distinct interfacial evolutions. As depicted in Figure 7g, the two systems exhibit markedly different sodium deposition kinetics. In the DME system, the number of deposited sodium atoms remains constant during the initial 5 ns, followed by a rapid increase. In contrast, the number in the EC system rises slowly and steadily from the beginning. This two-stage behavior in the DME system is corroborated by the analysis of the minimum sodium charge (Figure 7h). It is attributed to the properties of the anionic components: fluoride ions possess high electronegativity and ionic character, creating a deep potential well that traps $Na^+$ ions. Consequently, the NaF-dominated SEI initially acts as a reservoir where sodium undergoes a "solid solution" process; deposition of metallic sodium occurs only after the matrix reaches "solid solution saturation" (after ~5 ns).

In contrast, the mechanism in the $Na_2CO_3$-dominated SEI is fundamentally different. Due to the covalent nature of carbonate ions, the sodium species within the $Na_2CO_3$ matrix inherently carry a lower partial charge compared to the highly ionic NaF. Consequently, even a minor amount of sodium insertion is sufficient to induce a "metallic-like" character in localized regions of the SEI. This intrinsic electronic feature is the root cause of the system's high chemical reactivity: the ease with which the SEI matrix adopts metallic characteristics readily triggers the rapid decomposition of the electrolyte. As shown in Figure 7i and Supplementary Note S21, the EC system suffers from significantly higher sodium metal loss and electrolyte depletion compared to the DME system. Thus, despite the early onset of reduction behavior in the $Na_2CO_3$ system, the majority of the inserted sodium is consumed by side reactions rather than being stored. Consequently, this leads to a lower net deposition of metallic sodium compared to the NaF system. These findings reveal the atomic-scale nature of "sodium loss" during cycling: some $Na^+$ ions fail to deposit as metallic sodium after crossing the SEI. Instead, they are irreversibly "trapped" within the SEI layer because their binding energy with SEI components (such as defects) is stronger than that with bulk sodium. This process clearly demonstrates that interfacial chemistry directly governs sodium deposition morphology and Coulombic efficiency. Therefore, tuning the SEI composition is key to improving battery efficiency and cycling stability.

Beyond interfacial chemistry, the physical storage environment also dictates sodium storage morphology. For carbon-based electrodes, pore size drives a transition from an ionic adsorption state (in small pores) to a quasi-metallic deposition state (in larger pores), as shown in Supplementary Note S22. This indicates that optimizing carbon electrode pore structures to promote quasi-metallic formation is crucial for maximizing capacity. Crucially, this transition provides a physical basis for sodium loss: sodium stored in ionic states suffers from excessive binding energies due to strong substrate interactions, rendering it unable to participate in the electrochemical cycle.

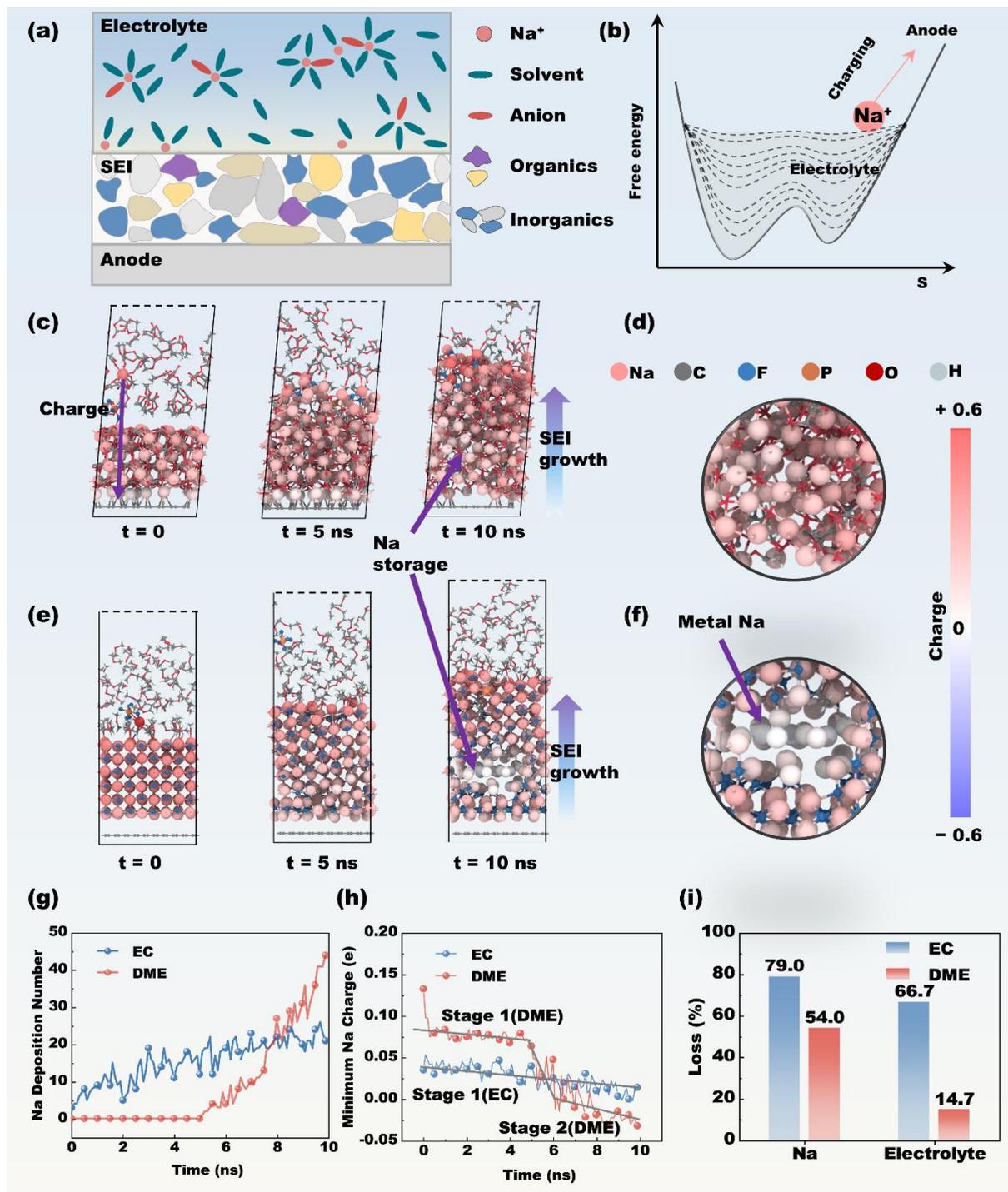

Figure 7. (a) Schematic of the anode interface model; (b) Metadynamics simulation schematic for the charging process; (c) Charging simulation snapshots (0–10 ns) in carbonate electrolyte with Na$_2$CO$_3$-dominated SEI; (d) Sodium clusters after charging in a Na$_2$CO$_3$ SEI; (e) Charging simulation snapshots (0–10 ns) in ether electrolyte with NaF-dominated SEI; (f) Sodium clusters after charging in a NaF SEI; evolution with different electrolytes (EC vs DME) of (g) sodium deposition number, (h) minimum sodium charge, and (i) sodium/electrolyte loss percentages. Atoms are repre-

sented in a ball-and-stick model (C: grey, F: blue, H: white, Na: pink, O: red, P: orange); Na atoms are colored according to their partial charges.

**Conclusions**

In this work, we develop an Interface Reactor sampling strategy combined with a charge-aware neuroevolution potential (qNEP) to simulate SEI formation and sodium storage at the Na-metal/electrolyte interface. This strategy overcomes critical limitations of conventional machine learning potentials by achieving stable, first-principles-accurate molecular dynamics simulations on the hundred-nanosecond scale. The simulations provide an atomistic picture of electrolyte-dependent SEI evolution. In carbonate electrolytes, rapid mixed co-formation produces a heterogeneous organic–inorganic interphase, including inorganic products (for example, $Na_2CO_3$), organic sodium salts and gaseous by-products (for example, $C_2H_4$ and CO). In ether electrolytes, the interphase evolves via surface-energy-controlled NaF crystallization, yielding ordered NaF nanocrystals that passivate the surface and suppress further solvent decomposition. These distinct interfacial chemistries reshape local solvation and govern Na plating. NaF-rich SEIs promote efficient metallic Na deposition with few parasitic reactions, whereas $Na_2CO_3$-rich matrices favour irreversible Na trapping and continued electrolyte consumption. More broadly, the Interface Reactor framework offers a general route to probe complex interfacial reaction networks in electrochemical systems.

**ACKNOWLEDGMENT**

The support of Shandong Province Excellent Youth Science Fund Project (2023HWYQ-022), Taishan Scholars Youth Expert Program of Shandong Province (tsqn202312002), Shenzhen Key Laboratory of Advanced Functional Carbon Materials Research and Comprehensive Application (No. ZDSYS20220527171407017) are acknowledged. The scientific calculations in this paper have been done on the HPC Cloud Platform of Shandong University.

Graphical Abstract

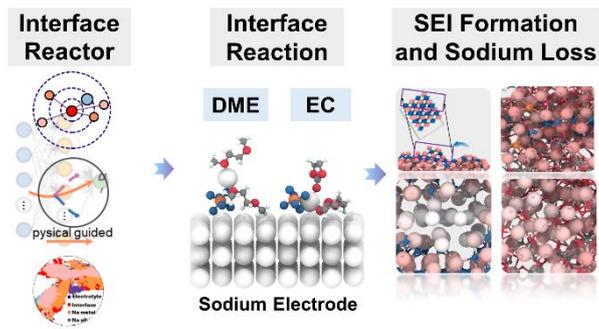